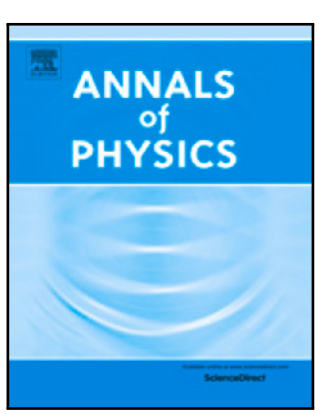

# Diffeomorphism-Invariant Quantities in Phase Space: More than Correlations

Álvaro Mozota Frauca

*Universitat Politècnica de Catalunya., Av. Diagonal, 649, Barcelona, Spain*

ABSTRACT

A popular view in the foundations of diffeomorphism-invariant theories is that their physical content is encoded in correlations or 'observables': a set of phase space functions that have vanishing Poisson brackets with the constraints related to diffeomorphisms. In this article I study the phase space structure of models with a temporal diffeomorphism invariance and prove a series of formal results that challenge this view in a few ways. First, I show how this view is not applicable to all the phase space trajectories of every diffeomorphism-invariant theory. Second, I show how correlations can be proved to be invariant only in a way that generalizes the standard definition of invariance and does not provide smooth phase space functions. Third, I prove that spatiotemporal relations are also invariant. Fourth, I prove that spatiotemporal structures are indispensable for defining the invariant content of diffeomorphism-invariant models. Finally, I comment that these results are expected to be generalizable for models invariant under $d$-dimensional diffeomorphisms, which represents a challenge for some views in the foundations of general relativity and quantum gravity.

## 1. Introduction

In the foundations and philosophy of physics there has been a great interest in studying the phase space structure of diffeomorphism-invariant theories and the way to define 'gauge-invariant quantities' in this class of theories. The reason is twofold. First, the understanding of general relativity and its diffeomorphism invariance has led to multiple conceptual issues and debates. [1] Second, the canonical quantization of the theory starts from its phase space formulation,[2] and it is therefore important to have control over the symmetries of the classical theory and their interpretation to safely proceed with its quantization.[3]

The predominant view in the quantum gravity literature[4] is that the physical content of classical diffeomorphism-invariant theories lies in its 'observables' $O$ which would be quantities satisfying an invariance condition, namely:

$$\{O, \mathcal{H}\} = 0\,, \tag{1}$$

*E-mail address:* alvaro.mozota@upc.edu.

[1] See the varied approaches and positions in [1–22].

[2] See the classical reviews [23,24] of the canonical quantization of gravity, Rovelli [5] for a book in which this quantization method is explained and put into practice, and Mozota Frauca [16] for a recent critical review of how this quantization process is affected by the diffeomorphism invariance of general relativity.

[3] In this article I will focus on the standard phase space formalization of theories like general relativity, which is the standard starting point for canonical approaches to quantum gravity. Notice, however, that there are other approaches to quantum gravity that make use of different formalisms and techniques. For instance, some recent works [25,26] propose to modify the phase space algebra of theories like general relativity and to study how this affects its quantization.

[4] The most influential author defending this view is Carlo Rovelli, who has defended this view since the 90s [2–4] until our days [5–7,27,28].

where $\mathcal{H}$ is a generator of diffeomorphisms in phase space. According to this view, these quantities would encode 'correlations', i.e., something like the value that some quantity takes when others take some given values. Based on this technical analysis of the gauge structure of theories like general relativity, many authors base their interpretation of the theory and justify the use of certain canonical quantization techniques to build quantum versions of diffeomorphism-invariant theories.

In this article I will analyze models with a temporal diffeomorphism invariance and I prove some results that challenge this view, which I call the 'observables' view or the 'frozen observables' view. At the end of the article I will comment that these results are expected to be generalizable to models with full spatiotemporal diffeomorphisms and to have consequences for both the interpretation of diffeomorphism-invariant theories and their quantization. However, in this article I will try to stick to proving the formal results and I will leave the analysis of its consequences to forthcoming work [29].

The main results I will prove in this article are:

1. There are systems for which correlations cannot be defined as phase-space functions satisfying $\{O, \mathcal{H}\} = 0$.
2. Even when we can define correlations as invariant quantities, these quantities satisfy a generalized invariance condition and are not, in general, smooth phase space functions.
3. Correlations are not the only invariant quantities. Spatiotemporal relations are also invariant quantities.
4. The underlying spatiotemporal structure of the theory is indispensable for defining its invariant quantities.

The structure of the article is the following. I start in Section 2 by introducing gauge theories in the Hamiltonian formalism, defining diffeomorphism-invariant models, and introducing and motivating the 'frozen observable' view. In Section 3 I introduce a few examples of models that contradict the 'observable' view, showing its limitations. Finally, in Section 4 I prove the 4 general claims above, each in a different subsection, and in Section 5 I briefly discuss them, their generalizability, and conclude.

Let me mention that some of the technical results in this article are directly related to some well-known examples and properties of temporal diffeomorphism invariant models.[5] For instance, in the examples in Section 3 systems appear that are periodic and non-monotonic in time. The quantum gravity community has discussed the use of this kind of system as a clock for some time now [30–32]. The general sentiment is that they constitute 'bad clocks' that shouldn't be used for building relational observables. However, some of my arguments in this article have to do with how we ought to interpret a model when no 'good clock' is in place. The frozen observables view needs 'good clocks' to make sense, and in general we cannot assume that we have them.

Similarly, the recent work [33] analyzes the observables of a system with a periodic clock and shows that the only relational observables one can define that are differentiable phase space functions that remain invariant along dynamical trajectories are defined only for degrees of freedom that evolve with the same periodicity as the clock. The discussion in that article does not analyze the consequences for the 'frozen observable' view, and my work can be seen as taking this kind of result one step further by performing such an analysis and obtaining more general results.

There is also a third relevant work that obtains significant results for the topic discussed in this paper. It is the article [34] where it is explicitly argued that observables for generic diffeomorphism invariant models, including general relativity, are not differentiable functions. In this work it isn't argued how this affects the 'frozen observable' doctrine and it does not consider whether correlations can be written as non-differentiable functions. My article complements this one by showing that correlations and temporal structures can be written as invariant quantities once the differentiability condition is lifted.

Finally, let me also comment that in this article I deal with the simple models and observables that are usually given in the foundations and philosophy of physics to discuss the interpretation of diffeomorphism-invariant models in general, but specifically general relativity and quantum gravity. Therefore, this article does not aim to provide a comprehensive analysis of the more recent, highly sophisticated approaches to observables that the quantum gravity community has developed in recent years—such as dressed observables, boundary reference frames, or covariant phase space constructions [35–38]. While an in-depth study of these modern technical developments is certainly interesting, the core of my argument does not require such sophisticated structures. Specifically, I prove that for simple mechanical models, invariant quantities are not necessarily differentiable functions, and that they include spatiotemporal relations. These results are expected to be extensible to theories with full four-dimensional diffeomorphism invariance, and are, hence, important for discussions in the foundations of general relativity and quantum gravity.

Furthermore, these findings are directly relevant also to physicists working within modern frameworks. My results show that observables shouldn't be expected to be smooth, differentiable functions. In particular, the toy model in this article highlights that even when dynamical systems are used as reference frames to single out points in spacetime, the relation between those spacetime points and the configurations of the dynamical system is not always smooth. This is especially relevant for modern relationalist approaches, and it still applies to covariant approaches that treat phase space not as a space of instantaneous states, but as the space of possible solutions to the equations of motion or as a space of boundary data. Even from this perspective, accounting for the non-differentiability of invariant quantities and acknowledging that spatiotemporal relations are themselves invariant remains an interesting contribution to our understanding of diffeomorphism-invariant theories.

A word on notation. To refer to points in phase space for generic systems I use the letter $x$ or the canonical coordinates $q^a, p_a$. For specific systems like those in Section 3 I adapt this notation to the system in play, e.g. $q_1, p_1, q_2, p_2, q_3, p_3$ and $q, p_q, t, p_t$. In Section 4.2 I split the coordinates of a point $x$ into a set $r$ referring to a reference system and those $(y)$ referring to the rest of the system. Lower-case letters $(q, t)$ refer to dynamical variables or coordinates, while upper-case $(Q, T)$ refer to fixed ones. $\tau$ is an arbitrary time parameter, while $t$ is a physically relevant time parameter (e.g., Newtonian time).

[5] I am thankful to an anonymous reviewer for bringing these results to my attention.

## 2. Diffeomorphism-invariant theories, 'observables', and correlations

### 2.1. Gauge transformations and invariant quantities

In the Hamiltonian formalism, gauge theories like electromagnetism are described by means of the constrained formalism.[6] This means that from the whole phase space only the points in a subspace of it, the constraint (hyper)surface, have the physical meaning of representing a possible instantaneous state of the gauge system. Multiple points on the constraint surface can represent the same state of affairs at a time. A gauge transformation is a transformation that maps a point on the constraint surface to another equivalent one. Gauge transformations (connected to the identity) are generated by the gauge generators $G$, which are linear combinations of the constraints $\phi_\alpha$ defining the constraint surface. This means that phase space coordinates $q^a, p_a$ transform according to the following equations:

$$\frac{dq^a(\xi)}{d\xi} = \{q^a, G\} \tag{2}$$

$$\frac{dp_a(\xi)}{d\xi} = \{p_a, G\}\,, \tag{3}$$

where $\xi$ is a continuous parameter that is associated with different transformations generated by the same generator $G$. When $\xi$ is 0 we get back $q^a, p_a$, i.e., the gauge transformation is just the identity. To obtain different gauge transformations one can choose different values of $\xi$ or pick different linear combinations of constraints to build a different generator $G$. If a phase space function $f$ is smooth, it will transform as:

$$\frac{df(\xi)}{d\xi} = \{f, G\} = \sum_a \frac{\partial f}{\partial q^a}\frac{\partial G}{\partial p_a} - \frac{\partial f}{\partial p_a}\frac{\partial G}{\partial q^a}\,. \tag{4}$$

We can define gauge-invariant quantities as those which do not change under gauge transformations. That is, $\frac{df(\xi)}{d\xi} = 0$ for all possible gauge generators. For smooth phase space functions this is equivalent to imposing:

$$\{f, \phi_\alpha\} = 0\,\forall\alpha\,. \tag{5}$$

That is, the smooth quantities that have vanishing Poisson brackets with all the constraints are invariant. For the 4-potential formulation of Maxwell's electromagnetism one can check that $\vec{\pi}$ and $\vec{\nabla}\times\vec{A}$, which correspond to the electric and magnetic fields respectively, are invariant phase space functionals at a time. These quantities are the physical fields in electromagnetism and are gauge independent.[7]

Given that this holds in electromagnetism, it is common to see this idea extended to any gauge theory and to diffeomorphism-invariant theories. That is, the physical content of a theory is identified with the phase space functions/functionals that have vanishing Poisson brackets with its constraints. However, when we apply this to the case of diffeomorphism-invariant theories in general, and general relativity in particular, we run into the technical and conceptual issues I discuss in this article.

### 2.2. Diffeomorphism-invariant theories

A diffeomorphism-invariant theory is one in which its solutions are parametrized by arbitrary spacetime coordinates that one can change freely (as long as one does it smoothly) to obtain an equivalent solution. This invariance is famously relevant because it lies at the core of general relativity and understanding it is crucial for its interpretation and quantization. For understanding this symmetry, it is customary to start by analyzing its one-dimensional version, just temporal diffeomorphisms, and then extrapolate to the full four-dimensional case. In this article, I will follow the same strategy and most of the discussion will turn around models with a temporal invariance, and I will leave the discussion of how the results found can be generalized to Section 5.

A temporal diffeomorphism-invariant theory is one in which its solutions are parametrized by an arbitrary time parameter or coordinate that one can change freely (as long as it is smooth) to obtain an equivalent solution. That is, given any solution $q^a(\tau)$, we can build an equivalent one by means of a diffeomorphism of the form $\tau' = \tau'(\tau)$ that makes us transform all the quantities accordingly. This means that if $q^a(\tau)$ is a scalar, it will transform into $q'^a(\tau'(\tau)) = q^a(\tau)$. For quantities of tensorial character like $\frac{dq}{d\tau}$ (or $g_{\mu\nu}$ in the field case) we need to apply the tensor transformation rules.

Given the symmetry of these theories, in the Hamiltonian formalism they are analogous to gauge theories. That is, we will find a constraint, the Hamiltonian constraint $\mathcal{H}$, which will play two roles: it will define the constraint surface of physically meaningful phase space points and it will be directly related to temporal diffeomorphisms. In some models, the constraint structure is slightly more complicated, as the Hamiltonian constraint is accompanied by another constraint. However, one can effectively treat this system as if there were only the Hamiltonian constraint associated with temporal diffeomorphisms.[8]

[6] Some of the standard references on the constrained formalism are [39–41].

[7] See [11,42–44] for detailed discussions of the constraint structure of electromagnetism.

[8] See [16] for a discussion of the constraint structure of diffeomorphism-invariant models and [42–45] for recent discussions on the role of the different constraints. For our discussion here, it won't be necessary to consider these subtleties and we can assume that any temporally diffeomorphism-invariant theory can be expressed as a constrained system with just one constraint.

For most of the discussion of this article I will discuss just temporally diffeomorphism-invariant models and leave the generalization to spatiotemporal diffeomorphism-invariant models to the discussion in Section 5, as from a conceptual and technical point of view it is enough with limiting to the simple case. I will also assume, for simplicity, that there is no other gauge symmetry involved. The generalization to a more general case is straightforward and does not provide any illuminating insight.[9] Therefore, I will consider systems with just one constraint, the Hamiltonian constraint $\mathcal{H}$.

The peculiarity of diffeomorphism-invariant theories is that the dynamics of these theories is formally equivalent to diffeomorphism transformations. The Hamiltonian of these systems is given by:

$$H_\lambda(x) = \lambda(\tau, x)\mathcal{H} \,. \tag{6}$$

That is, the Hamiltonian is proportional to the Hamiltonian constraint. I am using $x$ to refer to a point in phase space $\mathcal{P}$ and $\tau$ as the parameter of the evolution. $\lambda(\tau, x)$ is a function that can be chosen, reflecting the symmetry of the system. That is, for every choice of $\lambda$ we will have a different Hamiltonian, but all these Hamiltonians produce diffeomorphism equivalent trajectories in phase space. In particular, the dynamics is generated by Hamilton's equations of motion:

$$\frac{dx}{d\tau} = \{x, H_\lambda\} = \lambda\{x, \mathcal{H}\} \,. \tag{7}$$

Notice that as $\mathcal{H}$ is the only constraint, this equation is just the same as the equation for gauge transformations (4) when we apply that the generator of diffeomorphisms is $G = \mu\mathcal{H}$, for some suitably chosen function $\mu$. In this sense, for temporal diffeomorphism-invariant systems, the dynamical trajectories and gauge orbits coincide, as both are generated by the same constraint.

The solutions of these equations of motion for initial conditions $x_0$ can be written as:

$$x(\tau, x_0) = x(t_\lambda(\tau)) = (\exp t_\lambda\{\cdot, \mathcal{H}\})x_0 = \sum_{n=0}^{\infty} \frac{t_\lambda^n\{\cdot, \mathcal{H}\}^n}{n!} x_0 \,, \tag{8}$$

where $\{\cdot, \mathcal{H}\}^n x_0$ represents the application of $n$ times the Poisson bracket with $\mathcal{H}$ to $x_0$. That is, $\{\cdot, \mathcal{H}\}^0 x_0 = x_0, \{\cdot, \mathcal{H}\}^1 x_0 = \{x_0, \mathcal{H}\}, \{\cdot, \mathcal{H}\}^2 x_0 = \{\{x_0, \mathcal{H}\}, \mathcal{H}\}$, and so on. I have introduced the time $t_\lambda(\tau)$ as a convenient function, which is defined[10] as:

$$t_\lambda(\tau) = \int_0^\tau \lambda(\tau', x(\tau'))d\tau' \,. \tag{9}$$

This function is just $\tau$ if we set $\lambda = 1$. If not, it is diffeomorphically related to it. One can see it as a privileged way of parametrizing dynamical trajectories, as the solutions (8) look simpler when expressed just in terms of it. Indeed, for some systems it will correspond to some special temporal parametrization like Newtonian time, as we will see in the models of the next section.

Expressing solutions in terms of $t_\lambda$ also makes explicit that different choices of $\lambda$ correspond to different parametrizations of the same trajectory, i.e., to diffeomorphism transformed versions of the same curve in $\mathcal{P}$. To see this, notice for any $\lambda, \lambda'$ and $\tau$ there exists a value of $\tau'$ such that $t_\lambda(\tau) = t_{\lambda'}(\tau')$. This defines a diffeomorphism $\tau'(\tau)$ and given the form of the solution (8) we have that:

$$x'(\tau', x_0) = x(t_{\lambda'}(\tau'), x_0) = x(t_\lambda(\tau), x_0) = x(\tau, x_0) \,. \tag{10}$$

This is the condition for two trajectories in phase space to be diffeomorphism equivalent.

Let us now turn to gauge transformations. Given that these transformations are also generated by the Hamiltonian constraint, they have the following form:

$$x' = (\exp \sigma\{\cdot, \mathcal{H}\})x \,. \tag{11}$$

As the generator is the same as the generator of the dynamics, it is no surprise that the quantity $\sigma$ can be interpreted as the 'time' that we are shifting the point $x$. That is, $x'$ is just the same point we would have obtained if we had evolved the point $x$ following the equations of motion during a time $t_\lambda = \sigma$. In particular, we can see that replacing $x_0$ with $x_0'$ as the initial conditions of a dynamical trajectory has the effect of shifting the time of the whole evolution:

$$\begin{aligned} x(t_\lambda(\tau), x_0') = (\exp t_\lambda(\tau)\{\cdot, \mathcal{H}\})x_0' = (\exp t_\lambda(\tau)\{\cdot, \mathcal{H}\})(\exp \sigma\{\cdot, \mathcal{H}\})x_0 = \\ = (\exp (t_\lambda(\tau) + \sigma)\{\cdot, \mathcal{H}\}) = x(t_\lambda(\tau) + \sigma, x_0) \,. \end{aligned} \tag{12}$$

This means that if a point in the trajectory had a time (in the $t$ coordinatization) $t$, after applying the transformation to the initial conditions it will be the point at $t - \sigma$. This condition can be translated also as a transformation in $\tau$ coordinate according to the condition $t(\tau') = t(\tau) - \sigma$.[11] This corresponds to another diffeomorphism, i.e., a relationship $\tau'(\tau)$ between the coordinates of the trajectory taking $x_0$ as initial point and the coordinates that take $x_0'$ as initial points.

[9] For instance, we could consider the constraint structure of having a temporal diffeomorphism-invariant version of electromagnetism. Having both symmetries together wouldn't affect the analysis of either of them.

[10] Eqs. (8) and (9) together define $x(\tau, x_0), t_\lambda(\tau)$. The discussion in this article applies both for $x$-dependent and $x$-independent choices of $\lambda$ which means that $t_\lambda$ can depend on the initial conditions $x_0$. However, for simplicity I will write $t$ just as a $\tau$-dependent function. Considering explicitly the $x_0$ dependence only complicates the notation and does not affect the results in this article.

[11] This transformation depends on the choice of $\lambda$, and therefore it can encode some dependence on the phase space coordinates, even if I haven't written this dependency explicitly.

For the purposes of this article it will be enough to discuss gauge transformations as transformations acting at a moment of time or at the initial conditions of a trajectory. There is also a version of gauge transformations that acts on full trajectories and which is also equivalent to a diffeomorphism $\tau'(\tau)$. The details of this sort of transformation aren't needed for my discussion here, but notice that the distinction between the two kinds of transformation, local (in time) and global, is relevant in some discussions of gauge symmetries.

Finally, let me make a last remark about temporal diffeomorphism-invariant systems in the Hamiltonian formalism. Diffeomorphisms generated by the Hamiltonian constraint (i.e., transformations of the form (11)) do not cover the whole class of diffeomorphisms, as temporal reflections are left out. That is, transformations like $\tau'(\tau) = -\tau$ (or combinations of it with other diffeomorphisms) cannot be generated by the constraint. This means that there is a mismatch between diffeomorphisms in theories like general relativity and their formulation in the Hamiltonian formalism. In other words, using the Hamiltonian formalism imposes a temporal orientation to our models.

The relationship between diffeomorphism transformations and temporal evolution in temporally diffeomorphism-invariant models is what has given rise to a number of conceptual issues and to the kind of claims that I want to argue against in this article and which I will now introduce in more detail.

### 2.3. The 'frozen observables' view

Let me introduce a well-known example of temporal diffeomorphism-invariant model which has been used to motivate the 'frozen observable' view.[12] This is the parametrized version of the Newtonian dynamics of a single particle or body. In this model, instead of describing how the position $q$ of a particle evolves in Newtonian time $t$, we describe $q$ and $t$ as evolving with respect to an arbitrary parameter $\tau$. In other words, solutions to this model are trajectories $q(\tau), t(\tau)$, and we can smoothly transform $\tau$ into $\tau'$ to obtain an equivalent solution. The Hamiltonian (and only) constraint of this model is given by:

$$\mathcal{H}(q, p_q, t, p_t) = p_t + \frac{p_q^2}{2m} + V(q)\,, \tag{13}$$

where $p_q, p_t$ are the canonical momenta conjugate to $q, t$ and $V(q)$ is the potential energy of the particle. The transformations generated by the constraint in phase space transform a point $q, p_q, t, p_t$ into $q', p'_q, t', p'_t$ which is a point in the future according to the equations of motion of the theory.

How should one interpret $q', t'$? If we take that the constraint constitutes the Hamiltonian, then it is a different point from the original one and it represents how the system will be at some moment in the future. But if we take that the constraint generates instantaneous symmetry transformations just as in the case of a gauge theory, then we should interpret it as encoding the same physical meaning as $q, t$. In the observable view I am analyzing in this article, one takes the second option and claims that both points represent the same physics.

According to this view, if temporal evolution is just a gauge transformation, gauge-invariant quantities are necessarily constants of motion of the theory. For this reason they are sometimes called frozen observables. Luckily, for models like this one, one can build a family of quantities $Q_T(q, p_q, t, p_t)$ that satisfy the invariance condition $\{Q_T, \mathcal{H}\} = 0$ and out of which one can recover all the physics of the theory. These quantities are called correlations as they encode the value $Q$ that $q$ will take when $t$ takes the value $T$. The way to build them is to use the equations of motion of the theory to express $Q_T$ as a function of a state at a time $(q, p_q, t, p_t)$, which is taken as initial conditions. When we evolve $(q, p_q, t, p_t)$ we obtain a different point, but if we use the equations of motion of the theory to ask what will the position $Q$ of the particle be when the time is $T$, we will obtain the same answer as before, given that the point lies in the same dynamical trajectory. For this reason, the quantity $Q_T$ is a constant of motion and diffeomorphism invariant. Moreover, interpreted as a function of $T$, one can show that $Q_T$ satisfies Newton's equations of motion.

Given this successful example, the observable view is to extend it to any diffeomorphism-invariant theory. Therefore, it is based on the following postulates:

- **Formal definition of observable**: given any diffeomorphism-invariant theory, we define its observables $O$ to be the phase-space functions satisfying $\{O, \mathcal{H}_\mu\} = 0$ for every constraint $\mathcal{H}_\mu$ associated to diffeomorphisms.
- **Physical Interpretation**: the set of all (non-trivial) observables captures the physical content of the theory.
- **Correlations**: the correlations $Q_i(Q_1, Q_2, \ldots)$, which give the value of a physical quantity relative to the values of others, are the observables of diffeomorphism-invariant models and exhaust their physical content.

The notion of 'correlation' thus plays an important role in the 'observable' view and in my analysis in this article. Correlations like $Q_i(Q_1, Q_2, \ldots)$ are relational functions that give the value $Q_i$ that some target variable $q_i$ takes when a set of reference variables $q_1, q_2, \ldots$ take some values $Q_1, Q_2, \ldots$. For instance, in a model describing two particles, the correlation $X(Y)$ could represent the position of one of the particles relative to the position of the other one. Building correlations allows one to eliminate an explicit dependence on spacetime coordinates and introduce relational notions. The split between reference and target systems is usually taken to be a matter of convention.

[12] This example, its version as a pendulum (by choosing the right potential $V$), and relativistic versions of it have been extensively discussed in the literature around the 'observables' view [4–6,8,10–12,27,28,46–48].

Starting from these postulates, several authors have built their positions on different topics in the foundations of general relativity and they have justified different approaches to quantum gravity.[13] In particular, some of these authors take the 'observable' view to imply that time and spatiotemporal structures are eliminable from our diffeomorphism invariant theories.[14]

In this article I will prove some technical results that pose a challenge to these postulates and the positions based on them. Namely, I will show that while the 'observable' view works well for models like the one discussed above, it fails for general systems. I will show how for general systems one needs a more general definition of frozen observables for having correlations as invariant quantities. I will also show how spatiotemporal relations are also observables by the same standards, and that spatiotemporal structures are fundamental for developing these results, diminishing the claims that only correlations, and not spatiotemporal relations, are physical in diffeomorphism-invariant theories.

## 3. Models for which the definition of 'observable' fails

The definition of observable as a quantity that satisfies $\{O, \mathcal{H}\} = 0$ works well for models with monotonic variables, but for more general systems I will now show how this definition fails.

### 3.1. Two harmonic oscillators

Consider a diffeomorphism-invariant model of two harmonic oscillators. Its phase space is coordinatized by $q_1, p_1, q_2, p_2$ and its Hamiltonian constraint (the only constraint in the model) is:

$$\mathcal{H} = \frac{1}{2}\left(\frac{p_1^2}{m_1} + \frac{p_2^2}{m_2} + m_1\omega_1^2 q_1^2 + m_2\omega_2^2 q_2^2 - E\right), \tag{14}$$

where $m_i, \omega_i$ are positive magnitudes representing the masses and frequencies of the oscillators and $E$ is another positive quantity interpreted as the energy of the system. I will be assuming that the frequencies of the oscillators are not integer multiples of each other, i.e., that $\frac{\omega_1}{\omega_2} \notin \mathbb{Q}$. I will show how one can define correlations for this model.

Given a set of initial conditions $q_1^0, p_1^0, q_2^0, p_2^0$ the solutions of the equations of motion of the model can be expressed as:

$$q_i(\tau) = A_i(q_i^0, p_i^0)\sin(\omega_i t_\lambda(\tau) + \varphi_i(q_i^0, p_i^0)) \tag{15}$$

$$p_i(\tau) = m_i\omega_i A_i(q_i^0, p_i^0)\cos(\omega_i t_\lambda(\tau) + \varphi_i(q_i^0, p_i^0)). \tag{16}$$

Here I have introduced a few functions for convenience. $A_i(q_i^0, p_i^0) = \sqrt{(q_i^0)^2 + (\frac{p_i^0}{m_i\omega_i})^2}$ is the amplitude of each of the oscillators. $\varphi_i(q_i^0, p_i^0)$ is the initial phase of each of the oscillators which I have chosen to represent as an angle in the interval $[-\frac{\pi}{2}, \frac{3\pi}{2})$ and is given by:

$$\varphi_i(q_i^0, p_i^0) = \begin{cases} -\frac{\pi}{2} & \text{if } p_i^0 = 0, q_i^0 < 0 \\ \arctan\frac{m_i\omega_i q_i^0}{p_i^0} + \frac{\pi}{2}\left(1 - \frac{p_i^0}{|p_i^0|}\right) & \text{if } p_i^0 \neq 0 \\ \frac{\pi}{2} & \text{if } p_i^0 = 0, q_i^0 > 0 \end{cases} \tag{17}$$

This function is discontinuous for $p_i^0 = 0$ and $q_i^0$ negative as when we move from negative to positive $p_i^0$ we go from an angle of $\frac{3\pi}{2}$ to an angle of $-\frac{\pi}{2}$. This fact will be relevant later on.

The function $t_\lambda(\tau)$ in Eq. (15) is the one I defined in Eq. (9). Again, we see how it reflects the diffeomorphism invariance of the model, as different choices of $\lambda$ correspond to different diffeomorphism-related solutions $q_i(\tau), p_i(\tau)$.

Given all these definitions, the Eqns. (15) describe a set of two harmonic oscillators evolving with respect to an arbitrary parameter $\tau$, and they satisfy that $q_i(0) = q_i^0$ and $p_i(0) = p_i^0$.

From these solutions, if we want to follow the observable view, we would like to eliminate the arbitrary $\tau$ and build a correlation like $Q_2(Q_1, P_1)$, that is, the position that oscillator 2 takes when oscillator 1 takes position $Q_1$ and momentum $P_1$. However, it is immediate to see that as oscillator 1 is oscillating, just by giving its position and momentum we are not able to single out a unique moment of the trajectory, as there are infinitely many moments in which $q_1, p_1$ take those values. Therefore, as I am assuming that the frequencies of both oscillators are not tuned, we will have an infinite set of different correlations predicted by the theory $Q_{2,0}(Q_1, P_1), Q_{2,1}(Q_1, P_1), Q_{2,2}(Q_1, P_1)$...

Formally, to build the correlations we need to invert the relationship between $q_1, p_1$ and $t(\tau)$ so that we can rewrite $q_2(t(\tau))$ as a function $q_2(x_1, p_1)$. The problem is that, of course, the trigonometric equations $q_1(t) = q_1, p_1(t) = p_1$ have infinitely many solutions:

$$t(q_1, p_1, k) = \frac{1}{\omega_1}(\varphi_1(q_1, p_1) - \varphi_1(q_1^0, p_1^0) + 2\pi k) \quad k \in \mathbb{Z}, \tag{18}$$

[13] Again, see [2–7,27,28].

[14] Particularly telling is the title of the article [6]: 'Forget time'.

where $\varphi_1(q_1,p_1)$ is the same angle function (17) I defined for the initial conditions but applied to the position $q_1$ and momentum $p_1$. We therefore find an infinite sequence of times at which the oscillator is described by $q_1,p_1$. For each of these times we can express the position of the second oscillator as a function of $q_1,p_1$:

$$q_2(q_1,p_1,k) = A_2\sin(\frac{\omega_2}{\omega_1}\varphi_1(q_1,p_1) + (\varphi_2(q_2^0,p_2^0) - \frac{\omega_2}{\omega_1}\varphi_1(q_1^0,p_1^0)) + \frac{\omega_2}{\omega_1}2\pi k) \tag{19}$$

To build a 'frozen observable' we need to replace the variables $q_1$ and $p_1$ with some fixed values $Q_1$ and $P_1$ and the initial conditions $q_1^0,p_1^0,q_2^0,p_2^0$ by the phase space coordinates $q_1,p_1,q_2,p_2$. This defines the following functions:

$$\begin{aligned} &Q_2(Q_1,P_1,k,q_1,p_1,q_2,p_2) = \\ &\quad A_2(q_2,p_2)\sin(\frac{\omega_2}{\omega_1}\varphi_1(Q_1,P_1) + (\varphi_2(q_2,p_2) - \frac{\omega_2}{\omega_1}\varphi_1(q_1,p_1)) + \frac{\omega_2}{\omega_1}2\pi k) \end{aligned} \tag{20}$$

This quantity can be interpreted as the position that oscillator 2 takes when oscillator 1 is at $Q_1,P_1$ when oscillator 1 has done $k$ oscillations (starting from $q_1,p_1$). Now, do these functions satisfy the 'frozen observable' condition $\{Q_2,\mathcal{H}\} = 0$? One can show that $\{A_2,\mathcal{H}\} = 0$ and also that $\{\varphi_i(q_i,p_i),\mathcal{H}\} = \frac{\omega_i}{\lambda}$ as long as $q_i,p_i$ are not on the semi-line $p_i = 0, q_i < 0$. In these conditions, those two facts entail that $\{Q_2,\mathcal{H}\} = 0$. However, when $p_1 = 0$ and $q_1 < 0$, $\varphi(q_1,p_1)$ is a discontinuous function, and so is $Q_2$.[15] Therefore, we cannot define partial derivatives for it at this region of phase space and the Poisson bracket is not well-defined.

To see what is going on consider that we are following a trajectory generated by the Hamiltonian constraint. Let's start at the point $q_1 = -A_1, p_1 = 0$ and let's take $k = 0$. During the oscillation the quantity $Q_2(Q_1,P_1,0,q_1,p_1,q_2,p_2)$ remains constant and encodes the position that the oscillator 2 takes when the oscillator 1 is at $Q_1,P_1$ during that first oscillation. However, when we return to $q_1 = -A_1, p_1 = 0$ the argument in the sine abruptly changes as $\varphi_1(q_1,p_1)$ goes from $\frac{3\pi}{2}$ to $\frac{-\pi}{2}$, and we have that $Q_2$ also changes abruptly. At $q_1 = -A_1, p_1 = 0$ we are resetting the function $\varphi_1(q_1,p_1)$ to the value it had at the beginning of the oscillation, but $\varphi_2(q_2,p_2)$ is not reset, and hence it is no surprise that if we change the initial conditions for the second oscillator the value of $Q_2$ changes.

We can say more about the way it changes. What we are doing is replacing the original $\varphi_2(q_2,p_2)$ with the value that $\varphi_2$ takes after one oscillation of the first oscillator. In this sense, $q_2,p_2$ represents the state of the second oscillator at the beginning of the second oscillation. More generally, we can say that when we approach $q_1 = -A_1, p_1 = 0$ from the side of negative $p_1$ following a Hamiltonian trajectory the limit is not the function $Q_2(k)$ but the function $Q_2(k-1)$.

$$\lim_{\substack{q_1=-A_1\\ p_1\to 0^-}} Q_2(Q_1,P_1,k,q_1,p_1,q_2,p_2) = Q_2(Q_1,P_1,k-1,-A_1,0,q_2,p_2) \tag{21}$$

That is, the limit of the 'frozen observable' is telling us that if we want to compute the position of an oscillator in $k$ oscillations and we complete one oscillation, then we would have to use the observable corresponding to $k-1$ and not the one corresponding to $k$.[16]

If we want to build an observable that is truly constant along a Hamiltonian trajectory, we need to compensate for this effect. The way of doing so is to make $k$ not a constant but dependent on the trajectory in phase space. If $k(\tau)$ counts the number of oscillations of oscillator 1 from a given moment specified by $q_1,p_1,q_2,p_2$ to the desired moment in which we are interested, it should be of the form:

$$k(q_1,p_1,q_2,p_2) = k_0 - \int_0^{t(q_1,p_1,q_2,p_2)} \delta\left(\frac{p_1(t)}{m_1\omega_1 A_1}\right)(\frac{1-\frac{q_1}{|q_1|}}{2})dt \tag{22}$$

In the integral, $q_1(t),p_1(t),q_2(t),p_2(t)$ represents the trajectory (parametrized in Newtonian time, although any other choice is possible) between an initial point $q_1^0,p_1^0,q_2^0,p_2^0$ at $t=0$ and the point $q_1,p_1,q_2,p_2$.[17] The integral of the delta function counts how many times the trajectory between $q_1^0,p_1^0,q_2^0,p_2^0$ and $q_1,p_1,q_2,p_2$ crosses the semiline $p_1 = 0, q_1 = -A_1$. Therefore, if $k_0$ is the number of oscillations from the instant at which we had $q_1^0,p_1^0,q_2^0,p_2^0$ to the instant we are interested in, $k(q_1,p_1,q_2,p_2)$ counts the oscillations from the instant at which the state is $q_1,p_1,q_2,p_2$. As $q_1,p_1,q_2,p_2$ evolve, every time oscillator 1 completes an oscillation $k$ goes from $k$ to $k-1$, as one would have expected.

With this definition, we can build the observable:

$$Q_2(Q_1,P_1,k(q_1,p_1,q_2,p_2),q_1,p_1,q_2,p_2) \quad . \tag{23}$$

This represents the position that oscillator 2 takes when oscillator 1 is at $Q_1,P_1$ after $k$ oscillations from $q_1,p_1,q_2,p_2$, or, equivalently, after $k_0$ oscillations from $q_1^0,p_1^0,q_2^0,p_2^0$. This quantity is a constant of motion, as one can prove that now there is no problem at $p_1 = 0, q_1 = -A_i$ and the function remains the same.

Notice however that the function $k(q_1,p_1,q_2,p_2)$ is only defined for values of $q_1,p_1,q_2,p_2$ which lie in the same phase space trajectory as $q_1^0,p_1^0,q_2^0,p_2^0$. This is a subspace of dimension 1 in phase space. Therefore, $k(q_1,p_1,q_2,p_2)$ is not defined in the whole phase space and $Q_2$ either. For this reason, while we can claim that $Q_2$ is a constant of motion, we cannot claim that it satisfies $\{Q_2, \mathcal{H}\} = 0$.

[15] For the case of $p_2 = 0$ and $q_2 < 0$ there is no problem, as despite $\varphi_2$ not being continuous, its change of value just shifts the argument of the sine function in $Q_2$ by $2\pi$, which doesn't affect the sine value. Therefore, $Q_2$ is both continuous and differentiable when $p_2 = 0$ and $q_2 < 0$.

[16] This is just a consequence of $\lim_{\substack{q_1=-A_1\\ p_1\to 0^-}} 2\pi k - \varphi_1(q_1,p_1) = 2\pi k - \frac{3\pi}{2} = 2\pi(k-1) + \frac{\pi}{2} = 2\pi(k-1) + \varphi_1(-A_1,0)$.

[17] This is uniquely defined unless the frequencies of the oscillators are multiples of each other.

In this sense, this example shows how constants of motion can be built that represent the correlations that the observable view takes as physical. However, notice that there are important departures from that view. First, the observables found do not satisfy the condition $\{Q_2, \mathcal{H}\} = 0$ and they are defined just on a subspace of phase space. Second, the observables are indexed by $k$ or $k_0$ which means that there is some explicit reference to the temporal structure of the theory and that there is a natural temporal ordering of the observables. In this sense, the form of these observables shows that it is not just pure correlations that are constant, but that temporal order is also encoded in constant quantities.

A defender of the observable view could take a different position and claim that given that $q_1$ takes the same value an infinite number of times, what we should count as observable is not just one value of $q_2$, but the set of all values it takes. This could be defined as:

$$Q_2(Q_1, P_1, q_1, p_1, q_2, p_2) = \{Q_2(Q_1, P_1, k, q_1, p_1, q_2, p_2) \forall k \in \mathbb{Z}\} \quad , \tag{24}$$

or, equivalently:

$$Q_2(Q_1, P_1, q_1, p_1, q_2, p_2) = \{Q_2(Q_1, P_1, k(q_1, p_1, q_2, p_2), q_1, p_1, q_2, p_2) \forall k_0 \in \mathbb{Z}\} \quad . \tag{25}$$

The two definitions differ in that each takes one of the two versions of $Q_2$ I have introduced above, but they define the same set of values. This set can be said to be a constant of motion, as if we vary $q_1, p_1, q_2, p_2$ along a trajectory it remains the same set. However, there is no straightforward way of defining the partial derivative of a set, and the Poisson brackets remain ill-defined. In this sense, the definition of observable as quantity satisfying $\{O, \mathcal{H}\} = 0$ would still leave this set out.

The defender of the observable view can define this set and choose to ignore the temporal structure that one can define using constants of motion. However, there is a last fact that I want to emphasize: the temporal structure of the model has been indispensable for defining $Q_2$. That is, the search for a correlation of the form the value that $q$ takes when $y$ takes $Y$ doesn't proceed by looking for quantities that satisfy $\{O, \mathcal{H}\} = 0$, but, instead, it requires us to solve the equations of motion of the system and work on them so that we can express things conveniently.

### *3.2. Two harmonic oscillators as a clock*

The example above shows the limitations that we have when we use a non-monotonic variable as a clock to identify moments of time. Clearly, as there are multiple times at which the variable takes the same value, it fails to identify a single moment of time. Now I want to consider the case in which we have two non-monotonic variables to show how they can be used to identify moments of time.

Let's consider the same system of two harmonic oscillators together with a third system described by $q_3, p_3$ and non-interacting (for sake of simplicity) with the other two. The idea is to build observables of the form $Q_3(Q_1, P_1, Q_2, P_2, q_1, p_1, q_2, p_2, q_3, p_3)$. That is, to use the two harmonic oscillators as a reference system for the third system.

To the constraint and equations of motion in the previous subsection we would have to add the details of the third system, but with no loss of generality we will be able to write the solutions of its equations of motion as:

$$q_3(\tau) = q_3(t_\lambda(\tau), q_3^0, p_3^0) \quad . \tag{26}$$

Now, in order to express $q_3$ as a function of the state of the oscillators, we need to apply a very similar procedure to what we did in the previous example. As the equations of motion for the two oscillators do not change, we can invert the relationships between the oscillators' positions and momenta and time just as in the previous section:

$$t(q_1, p_1, k_1) = \frac{1}{\omega_1}(\varphi_1(q_1, p_1) - \varphi_1(q_1^0, p_1^0) + 2\pi k_1) \quad k_1 \in \mathbb{Z} \tag{27}$$

$$t(q_2, p_2, k_2) = \frac{1}{\omega_2}(\varphi_2(q_2, p_2) - \varphi_2(q_2^0, p_2^0) + 2\pi k_2) \quad k_2 \in \mathbb{Z} \tag{28}$$

As in the previous example, we can express the time as a function of the position and momentum of one of the oscillators and the number of oscillations it has completed. However, in this case we can formally eliminate the reference to the number of oscillations by expressing $k_1$ and $k_2$ as a function of $q_1, p_1, q_2, p_2$. For doing so, we impose that $t(q_1, p_1, k_1) = t(q_2, p_2, k_2)$, and this leads to the following equation for $k_1, k_2$:

$$k_1 - \frac{\omega_1}{\omega_2} k_2 = \frac{1}{2\pi}\left(\frac{\omega_1}{\omega_2}(\varphi_2(q_2, p_2) - \varphi_2(q_2^0, p_2^0)) - (\varphi_1(q_1, p_1) - \varphi_1(q_1^0, p_1^0))\right) \tag{29}$$

This equation implicitly defines $k_1$ and $k_2$ as functions of $q_1, p_1, q_2, p_2$. If $k_1$ and $k_2$ were real numbers, this equation would have infinite solutions, but as they are integers, the properties of this equation are rather peculiar. For most of phase space, this equation does not have any integer solution. We find solutions to this equation when $q_1, p_1, q_2, p_2$ is a point in the trajectory of the equations of motion for the system with initial conditions $q_1^0, p_1^0, q_2^0, p_2^0$.[18] In particular, if we set $q_1 = q_1^0, p_1 = p_1^0, q_2 = q_2^0, p_2 = p_2^0$ we obtain the equation $k_1 - \frac{\omega_1}{\omega_2} k_2 = 0$, which for $\frac{\omega_1}{\omega_2} \notin \mathbb{Q}$ only has as a solution the expected $k_1 = k_2 = 0$.

[18] Similarly, as in the equation the amplitudes $A_1$, $A_2$ do not play any role in the equation, if a point $q_1, p_1, q_2, p_2$ has solutions $k_1, k_2$, a point related by an amplitude transformation ($\alpha q_1, \alpha p_1, \beta q_2, \beta p_2$, with $\alpha, \beta > 0$) has the same solutions.

A way of seeing these properties of Eq. (29), is by thinking of it as equivalent to expressing the real number on the right-hand side as a linear combination of 1 and $\frac{\omega_1}{\omega_2}$ with integer coefficients. The real numbers are an infinite-dimensional vector space over the rationals. 1 and $\frac{\omega_1}{\omega_2}$ are a basis for a subspace of dimension 2, which means that most of the real numbers cannot be expressed in the form $k_1 \cdot 1 + k_2 \cdot \frac{\omega_1}{\omega_2}$, and for the ones in this subspace, the decomposition is unique.

Having said this, we can treat $k_1$ (the same holds for $k_2$) as a function $k_1(q_1, p_1, q_2, p_2)$ and write the expression for time in a way that is not explicitly $k$-dependent:

$$t(q_1, p_1, k_1(q_1, p_1, q_2, p_2)) = \frac{1}{\omega_1}(\varphi_1(q_1, p_1) - \varphi_1(q_1^0, p_1^0) + 2\pi k_1(q_1, p_1, q_2, p_2)) \quad . \tag{30}$$

Now we can finally express $q_3(\tau)$ with no explicit time dependence:

$$q_3(\tau) = q_3(t(q_1, p_1, q_2, p_2), q_3^0, p_3^0) \tag{31}$$

By repeating the trick in the previous section, we can convert this phase-space function in a 'frozen observable' by converting the variables $q_1, p_1$, etc. into fixed values and the initial conditions $q_1^0, p_1^0$, etc. into phase-space variables. That is, we can define the quantity:

$$Q_3(Q_1, P_1, Q_2, P_2, q_1, p_1, q_2, p_2, q_3, p_3) \quad , \tag{32}$$

In other words, we can express the state that system 3 will have when the oscillators are at $Q_1, P_1, Q_2, P_2$ as a function of a set of initial conditions $q_1, p_1, q_2, p_2, q_3, p_3$. From an intuitive point of view, this result has to be independent of the choice of initial conditions, and therefore constant along a Hamiltonian trajectory.

Clearly, as $k_1(q_1, p_1, q_2, p_2, q_1^0, p_1^0, q_2^0, p_2^0)$ is not defined in the whole phase space, when we change the roles and obtain $k_1(Q_1, P_1, Q_2, P_2, q_1, p_1, q_2, p_2)$ we will find a similar situation. That is, $k_1$ is only well-defined for $q_1, p_1, q_2, p_2$ which correspond to initial conditions for $Q_1, P_1, Q_2, P_2$. In this sense, one can simply not define partial derivatives of this function, nor a Poisson bracket with the Hamiltonian constraint.

Nevertheless, one can check that the quantity (32) is constant along Hamiltonian trajectories. Just as in the previous example, one can show that when an oscillator approaches the discontinuity in $q_i = -A_i, p_i = 0$, the value of $k_i$ goes from $k_i$ to $k_i + 1$, and this implies that (32) remains constant along a trajectory.

## 4. General results

The analysis of the models in the previous section will be the starting point for my general argument for the main claims of this article:

1. There are systems for which correlations cannot be defined as phase-space functions satisfying $\{O, \mathcal{H}\} = 0$.
2. Even when we can define correlations as invariant quantities, these quantities satisfy a generalized invariance condition and are not, in general, smooth phase space functions.
3. Correlations are not the only invariant quantities. Spatiotemporal relations are also invariant quantities.
4. The underlying spatiotemporal structure of the theory is indispensable for defining its invariant quantities.

These four claims go against the core of the frozen observable view. Claims 1 and 2 suggest that the frozen observable view should consider a generalized definition of observable that allows for non-differentiable functions. Claims 3 and 4 support the claim that spatiotemporal structure and relations should be considered observable and physical just as correlations.

I argue for each of them in separate subsections.

### *4.1. The definition of correlation as 'observable' does not apply to every system*

The models discussed in the previous section are examples of cases for which the claim that correlations are phase space-functions satisfying $\{O, \mathcal{H}\} = 0$ does not apply. In both cases I considered all the possible ways of defining correlations, and none of them was of the form of a phase space function satisfying $\{O, \mathcal{H}\} = 0$. In this sense, the first claim is proven, which means that the 'observable' view of diffeomorphism-invariant theories starts from a limited technical definition which is unable to represent correlations as observables in every dynamical model.

In particular, the examples show two sources of trouble for the definition of correlations in terms of 'observables'. Both have to do with the way we identify moments of time by using some of the phase space variables. The first issue has to do with using a clock that is not monotonous in time, which means that there is more than one instant of time at which the clock is in the same state, rendering the clock unable to tell apart different instants of the evolution. Formally, we say that the temporal evolution of the clock is not injective.

**Definition 4.1** (*Injective Temporal Evolution*)**.** We say that the temporal evolution of a system or subsystem described by a trajectory on phase space or a subspace of it $x(\tau)$ is injective if $x(\tau) = x(\tau') \implies \tau = \tau'$. That is, that the trajectory does not go through the same point more than once.

When the temporal evolution of the clock is not injective, one would have to introduce some reference to which of the times the clock is at a given state in order to be able to tell apart different instants of the evolution. It is for this reason that we could only define correlations for the first model that make reference to the number of oscillations $k$. If we wanted to get rid of this, we could define sets of correlations. In either case, we found that the observables aren't well-defined continuous and differentiable phase-space functions, and that they do not satisfy $\{O, \mathcal{H}\} = 0$.

The second model exemplifies a different kind of issue. The evolution of the set of two clocks is injective: for any state $(q_1, p_1, q_2, p_2)$ along a trajectory $(q_1, p_1, q_2, p_2)(t)$ there is only one value of $t$ at which the trajectory is at $(q_1, p_1, q_2, p_2)$. This allows inverting the relation and defining a function $t(q_1, p_1, q_2, p_2)$ (Eq. (30)) which gives us the time at which a certain state obtains in a given trajectory. However, this function cannot be interpreted as a function in phase space $\mathcal{P}$ but just as a function on the trajectory. Indeed, recall that in the definition (30) there appears $k_1(q_1, p_1, q_2, p_2)$, which is defined only for points on the trajectory. When building the correlation $Q_3(Q_1, P_1, Q_2, P_2, q_1, p_1, q_2, p_2, q_3, p_3)$ I made use of the time function, and hence it is a function which is defined only on the trajectory. One cannot compute $\{Q_3, \mathcal{H}\}$, and hence, the definition of observable does not apply.

### *4.2. Correlations as diffeomorphism-invariant quantities*

The discussion above invites us to find correlations not in quantities that satisfy $\{O, \mathcal{H}\} = 0$ but in quantities that may not be defined as phase space functions, given that they will be defined only along any given trajectory. These quantities will be well-defined as long as we use the appropriate reference subsystem, i.e. clocks. That is, they will be well-defined as long as the evolution of the reference subsystem is injective.

In this section, let me call $r$ all the phase space coordinates corresponding to the reference subsystem and $y$ the coordinates corresponding to the rest of the system. That is, $x = (r, y)$. The correlations we are looking for are functions $Y(R, r, y)$, where the capitalized $R$, $Y$ represent fixed values and $r$, $y$ phase space coordinates that are allowed to vary. $r$, $y$ play the role of initial conditions.

**Theorem 1.** *Let $\mathcal{P}$ be a phase space with coordinates $x = (r, y)$, where $r$ are the phase-space coordinates of a reference subsystem and $y$, the phase-space coordinates of the rest of the system. Let $\mathcal{H}$ be the Hamiltonian constraint that generates the dynamical evolution of the system.*

*Assume that $\Gamma \subset \mathcal{P}$ is a phase-space trajectory generated by $\mathcal{H}$ such that the dynamics of the reference system is injective during this trajectory.*

*Then, one can define the correlation $Y(R, r, y)$, for $(r, y) \in \Gamma$, which represents the value that $y$ takes when $r = R$ on the trajectory $\Gamma$. $Y(R, r, y)$ is constant along $\Gamma$.*

**Proof.** Given a choice of Hamiltonian, the dynamics defines the temporal evolution of any phase space function, as given by (8). In particular, it defines $r(t_\lambda(\tau), r_0, y_0)$ and $y(t_\lambda(\tau), r_0, y_0)$.

By assumption, the dynamics of the reference system is injective. This means that one can invert the relation between $r$ and $\tau$ and define $\tau$ as a function of $r$:

$$\begin{aligned} \tau_\lambda : \mathcal{R} \times \mathcal{P} &\longrightarrow \mathbb{R} \\ (r, r_0, y_0) &\longmapsto \tau_\lambda(r, r_0, y_0) \end{aligned} \tag{33}$$

The domain for the $r$ argument, $\mathcal{R}$, is not the full subspace of phase space corresponding to the reference subsystem, but just the subspace of values of $r$ along the trajectory corresponding to the initial values $r_0, y_0$.

We can define the correlations by composing the two functions:

$$y(r, r_0, y_0) = y(t_\lambda(\tau_\lambda(r, r_0, y_0)), r_0, y_0)\,. \tag{34}$$

This represents the state of the subsystem $y$ when the reference subsystem is at the state $r$ as a function of the initial conditions $r_0, y_0$. It is independent of the choice of Hamiltonian, and hence of parametrization. This is because, for fixed initial conditions, two parametrizations corresponding to different choices $\lambda$ satisfy that if $x'(\tau') = x(\tau)$, then $t_{\lambda'}(\tau') = t_\lambda(\tau)$. This makes $t_\lambda(\tau_\lambda(r, r_0, y_0))$ $\lambda$-independent, and $y(r, r_0, y_0)$ too. Therefore, we can define the function:

$$t(r, r_0, y_0) = t_\lambda(\tau_\lambda(r, r_0, y_0))\,, \tag{35}$$

which is independent of $\lambda$. Then, we can write $y(r, r_0, y_0)$ simply as:

$$y(r, r_0, y_0) = y(t(r, r_0, y_0), r_0, y_0)\,. \tag{36}$$

Now, we can rename our variables to obtain a correlation of the expected form:

$$Y(R, r, y) = Y(t(R, r, y), r, y)\,. \tag{37}$$

As before, by the capitalized $R, Y$ we represent fixed values and $r, y$ act as initial conditions. Let us now check that these are constant when we vary $r, y$ along a dynamical trajectory (equivalently, that they are unchanged under a transformation generated by the Hamiltonian constraint). We need to evaluate the function $Y(R, r, y)$ when we replace $r, y$ with $r', y'$ related by a gauge transformation. Following the discussion in Section 2.2 and applying Eq. (12), we have that:

$$Y(t(R, r', y'), r', y') = Y(t(R, r', y') + \sigma, r, y)\,. \tag{38}$$

As discussed in Section 2.2, when the initial conditions are evolved, $t$ changes accordingly following the relation $t(R, r', y') = t(R, r, y) - \sigma$. Then, we have that:

$$Y(t(R, r', y'), r', y') = Y(t(R, r', y') + \sigma, r, y) = Y(t(R, r, y), r, y). \tag{39}$$

Therefore, $Y(R, r, y)$ does not change under the application of transformations generated by $\mathcal{H}$ to $r, y$. In other words, it remains constant along the dynamical trajectory $\Gamma$. □

I have just shown how for any general dynamical system one can define correlations as functions that are constant along a trajectory as long as one has a subsystem with an injective evolution. These quantities are just the formalization of the fact that given any deterministic dynamical system, one can choose any point along a dynamical trajectory as a set of initial conditions to determine a correlation and it is of course independent of this choice. These functions are not, in general, defined on the whole phase space, but just along trajectories $\Gamma$. This implies that for the quantities $Y(R, r, y)$ one cannot define the Poisson bracket and that therefore they do not satisfy a relation of the form $\{Y(R, r, y), \mathcal{H}\} = 0$.

The example in the previous section illustrates these properties of the correlations, and in particular, that they are not functions defined on the whole phase space. Let me highlight that whether a subsystem can be used as a reference subsystem depends on the given trajectory or, equivalently, on the initial conditions. For instance, in the example with the two harmonic oscillators, if we had one of the oscillators standing still at the equilibrium position, then the subsystem of the two oscillators wouldn't be a good reference system, as its trajectory in its subspace of phase space wouldn't be injective.

For systems with no injective subsystems, one can only define correlations in a more careful manner. One way to do it is to restrict the range of evolution to a range in which the reference subsystem is injective. For instance, in our model of the two harmonic oscillators, if we consider just one period of oscillation, the correlation (20) for $k = 0$ would capture this correlation and would be constant as long as we stick to that first oscillation. Similarly, one could build correlations that referred not to the next time $R$ obtains but to the $k$th time it does, just as we did for the double harmonic oscillator model. These quantities could be defined to be constant during parts of the trajectory at which the evolution of $r$ is injective. To have them as truly constant one would have to define $k$ not as a constant but as a dynamical quantity, as we did in (23). However, this would not be trivial and it would still encode some temporal structure to tell apart the different moments at which the reference system takes the value $R$. Notice that in this case it would still be true that these correlations would generally be defined only on the trajectories and not on the whole phase space.

Alternatively, one could define correlations as sets. One could define the set of all possible correlations by building the correlations for each segment of the evolution for which the reference subsystem is injective and collecting them. Another way of defining this set is more analogous to the way I have defined correlations for temporal evolutions of systems in which a subsystem is injective. We start with the definition of temporal evolution as given by Eq. (8) once we choose a Hamiltonian. Now we cannot define a function $\tau$ as we did in (33), as $r(\tau)$ is not invertible, but we can define an analogous set-valued function:

$$\begin{aligned} T : \mathcal{R} \times \mathcal{P} &\longrightarrow P(\mathbb{R}) \\ (R, r, y) &\longmapsto T(R, r, y) = \{\tau \in \mathbb{R} | r(\tau, r, y) = R\} \end{aligned} \tag{40}$$

That is, we define $T$ as the set of all values of $\tau$ at which $r(\tau, r, y)$ takes a given value $R$. $P(\mathbb{R})$ is the power set of $\mathbb{R}$, i.e., the set of all possible subsets of $\mathbb{R}$. With this, we can define the correlations as the set:

$$Y(R, r, y) = \{y(\tau, r, y) | \forall \tau \in T(R, r, y)\} \tag{41}$$

Similar to the injective case, this set doesn't depend on the temporal parametrization chosen and is constant along dynamical trajectories. As argued above, the functions involved in this definition are such that one cannot in general claim that $\{Y, \mathcal{H}\} = 0$. Besides the difficulties in defining the Poisson bracket for a set-valued function, the argument of the function is not the whole phase space, but just a dynamical trajectory.

### *4.3. Temporal structures as diffeomorphism-invariant quantities*

The previous discussion has shown how the expectation that correlations of the form $Y(R)$ are quantities satisfying $\{Y(R), \mathcal{H}\}$ was wrong, as correlations can be defined but in a more general form that does not always allow for the definition of Poisson brackets. The good news for the 'observable' view is that I have shown that correlations can be defined as constants of motion, just as they claimed. However, in this section I will prove that it is not only correlations that can be defined in this way, but that temporal structures can be defined in this way too. In this sense, the argument that only correlations are physical can be challenged.

As before, I will start by restricting the discussion to systems with injective evolution. As we are not interested in correlations anymore, in this section I turn to the notation $x$ to represent points in the whole phase space.

**Theorem 2.** *Let $\mathcal{P}$ be a phase space with coordinates $x$ and let $\mathcal{H}$ be the Hamiltonian constraint that generates the dynamical evolution of the system.*

*Assume that $\Gamma \subset \mathcal{P}$ is a phase-space trajectory generated by $\mathcal{H}$ such that the dynamics is injective during this trajectory. Assume that $X_1$, $X_2 \in \Gamma$ are two fixed points on the trajectory.*

*Then, one can define the temporal order function* TO$(X_1, X_2, x)$ *evaluated at* $x \in \Gamma$ *such that*

$$\mathrm{TO}(X_1, X_2, x) = \begin{cases} 1 & \textit{if } X_1 \textit{ happens after } X_2 \\ -1 & \textit{if } X_1 \textit{ happens before } X_2. \\ 0 & \textit{if } X_1 = X_2 \end{cases} \tag{42}$$

*This function remains constant if we vary* $x$ *along* $\Gamma$*. 'Before' and 'after' are defined according to the time orientation of the flow of the Hamiltonian formalism.*

The idea of this theorem is that one can define a 'frozen observable' that does not correspond to a correlation but with the temporal ordering between two events.

**Proof.** The proof is analogous to the one of Theorem 1. As the system is by assumption injective, given any choice of $\lambda$ we can invert the temporal evolution to define $\tau_\lambda(x, x_0)$:

$$\begin{aligned} \tau_\lambda : \mathcal{X} \times \mathcal{P} &\longrightarrow \mathbb{R} \\ (x, x_0) &\longmapsto \tau(x, x_0) \end{aligned} \tag{43}$$

This is analogous to Eq. (33) with the modification that now the first argument is a complete set of phase space coordinates and its associated domain $\mathcal{X}$ is the set of all phase space points on the time evolution with $x_0$ as initial conditions. Given any two points $X_1, X_2$ in the evolution, for initial conditions $x$ we can compute the quantity:

$$\tau_\lambda(X_1, x) - \tau_\lambda(X_2, x) \tag{44}$$

This is the difference in time $\tau$ in between the events at which $X_1$ and $X_2$ obtain. This quantity depends on the choice of $\lambda$ and on the choice of initial conditions, but its sign doesn't, and therefore we can define the time ordering function as:

$$\mathrm{TO}(X_1, X_2, x) = \mathrm{sign}(\tau_\lambda(X_1, x) - \tau_\lambda(X_2, x)) \,. \tag{45}$$

This function gives 1 when $X_1$ happens after $X_2$ in the evolution generated by $\lambda$, $-1$ if it happens before, and 0 if $X_1 = X_2$. As argued above, changing $\lambda$ or $x$ along a time evolution corresponds to a diffeomorphism connected to the identity,[19] and it cannot change temporal orientation. Therefore, if for a choice of $\lambda$ and $x$, $X_1$ happens before $X_2$, this will happen for any choice. This means that the definition of the time orientation is parametrization independent and constant if we evolve $x$ along a dynamical trajectory. □

The function just defined has the property of being a 'frozen observable' in the sense of being a quantity that does not change along a temporal evolution. Again, contrary to the correlations view, it is a quantity that goes beyond a mere correlation between the subsystems at an instant of the evolution. And just as the correlations discussed in the previous section, in general it is not a function defined on the whole phase space, but on a subspace of it. This means that for these quantities we cannot in general define Poisson brackets and that they do not satisfy a condition of the form $\{O, \mathcal{H}\} = 0$.

One can define further temporal structures as 'frozen observables'. For instance, one can define an in-betweenness function:

$$\mathrm{IB}(X_1, X_2, X_3, x) = \mathrm{TO}(X_1, X_2, x)\,\mathrm{TO}(X_2, X_3, x) = \begin{cases} 1 & \text{if } X_2 \text{ happens in between } X_1 \text{ and } X_3 \\ -1 & \text{if } X_2 \text{ happens before or after both } X_1 \text{ and } X_3\,. \\ 0 & \text{if } X_1 = X_2 \text{ or } X_2 = X_3 \end{cases} \tag{46}$$

This function takes three instants in the evolution and tells us whether the second argument happens in between the two others, before or after both of them, or is one of the two. Like the time order function, it is a function of $x$ as a set of initial conditions but it is defined only on the trajectory on which $X_1, X_2$, and $X_3$ lie (these three need to be part of the same trajectory), and it is constant when we change $x$ along a dynamical trajectory. The in-betweenness relationship has a virtue that temporal order lacks, it is invariant also under changes of temporal orientation, and hence of the full diffeomorphism group. In the Hamiltonian formalism these changes aren't very relevant, but for the general discussion of diffeomorphisms and invariant quantities, some authors take in-betweenness relations to be fundamental and invariant.

Temporal relationships usually include a metric component, that is, duration. Diffeomorphism-invariant models can be interpreted as models in which duration plays no role as temporal parametrization is arbitrary. However, in many models one can define a privileged parametrization that corresponds to duration. In particular, in my discussion I have used the time parameter $t$, which is a useful parametrization that turns out to be the Newtonian parametrization for the models discussed in Section 3. This allows us to define the duration between two events as:

$$D(X_1, X_2, x) = t(X_1, x) - t(X_2, x) \,. \tag{47}$$

As the quantities defined above, this is invariant under changes of $\lambda$ and initial conditions $x$. That is, it would be another 'frozen observable' which in general wouldn't satisfy the condition $\{O, \mathcal{H}\}$.

[19] More precisely, these diffeomorphisms were defined by the conditions $t_\lambda(\tau) = t_{\lambda'}(\tau')$ for a change of $\lambda$ and $t_\lambda(\tau') = t_\lambda(\tau) - \sigma$ for a shift in initial conditions. As $t_\lambda(\tau)$ is an increasing function, $\tau(\tau')$ also is, and, therefore, orientation is preserved.

The dynamics of Hamiltonian models implies that any non-injective model is periodic, that is, if a trajectory in phase space visits the same point twice, then it has to repeat itself. For these models, observables like temporal ordering or in-betweenness make little sense, and duration could be defined by restricting the definition to just a period of the evolution.

In spacetime theories one can further generalize these definitions to represent the more diverse set of spatiotemporal relations. For instance, in a temporal diffeomorphism-invariant version of field dynamics in Minkowski spacetime, one could define functions representing whether two points are timelike, spacelike or lightlike separated or the proper time along a geodesic joining two timelike points. I will discuss the generalization to models with full spacetime diffeomorphism invariance in Section 5.

The analysis in this subsection has shown how the claim that 'correlations are the gauge-invariant quantities of diffeomorphism-invariant theories' is not true, as temporal or spatiotemporal relations can also be expressed in a way that is independent of the point $x$ along a trajectory which is used as initial conditions. Therefore, temporal relations belong to the 'frozen observables' category just as correlations do. This goes against the 'frozen observable' claim that the observables, in the technical and intuitive senses, of diffeomorphism-invariant models are just the correlations they predict.

### *4.4. The role of temporal structure*

Finally, in this last section I want to highlight that in order to derive all the claims in this article, the spatiotemporal structure of the models has played an indispensable role. That is, even if the 'observable view' claims that the physics lies just in correlations between physical quantities, that time is an eliminable concept from our theories and that finding the content of the theory consists of finding quantities that satisfy $\{O, \mathcal{H}\} = 0$, the truth is that my analysis of the symmetry in the Hamiltonian formalism takes time and dynamics to be fundamental. The jump that the 'observable view' does from claims about a simple model to more general theories, including general relativity, just does not follow.

To find the physical content of a model invariant under temporal diffeomorphisms, one does not have to find quantities that satisfy $\{O, \mathcal{H}\} = 0$, but to solve its equations of motion and analyze trajectories in phase space, which naturally represent a succession of states of a system. In other words, the dynamics of these models fundamentally makes use of temporal structures, and the analysis of the symmetry does not eliminate the role of these structures. There is no reason to believe that this would be different in the case of spacetime theories like general relativity. In this sense, the results in this article give the formal reasons to reject the physical interpretation of diffeomorphism-invariant models that the 'observable view' defends.

## 5. Discussion

The four results in the previous section have been proven in the context of theories with a temporal diffeomorphism invariance, but one can expect them to be readily generalizable to spacetime theories with full 4-dimensional diffeomorphism invariance. One would need to deal with a larger symmetry group, but definitions such as the definition of injective evolution can be straightforwardly adapted so that one can define injective fields or collections of fields if they satisfy that $\varphi(x^\mu) = \varphi(y^\mu)$ implies $x^\mu = y^\mu$. Therefore, it seems that there is no obstacle to defining correlations and spatiotemporal relationships as 'frozen observables', but, just as in the one-dimensional case, they wouldn't necessarily be, and in general they won't be, quantities satisfying $\{O, \mathcal{H}_\mu\} = 0$. Therefore, the challenge to the observable view extends to the four-dimensional case.

At a more technical level, one could worry that as the constraint structure of diffeomorphism-invariant models in higher dimensions, most relevantly, general relativity, is more complicated than the single constraint structure of the models described in this article, some technical issue blocks the applicability of my results to a more general case. However, as far as I know, the constraint structure of this kind of theory is well-understood, and it is clear that the gauge generators in phase space generate transformations that correspond to the phase-space version of diffeomorphisms. In this sense, I believe that my results are generalizable, even if, of course, this is something that deserves a more careful and detailed study and is left for future work.

The analysis of the diffeomorphism invariance of general relativity, in the Hamiltonian formalism or otherwise, is fundamental for our understanding of the theory and its interpretation. Moreover, it also plays a role in the justification and interpretation of a number of approaches to quantum gravity, and hence it deserves careful study. The 'frozen observable' view is certainly influential and is extended in the quantum gravity community. However, the results in this article contradict the technical claims in which it is based and it is therefore weakened. To insist, the results in this article suggest a revision of the definition of observable and to consider that spatiotemporal relations are also observable in the intuitive sense of the term, and, hence, physical. This has ramifying consequences in the foundations of general relativity and quantum gravity that will be explored in future work.

The results in this paper were framed within the canonical formalism, but they are relevant for approaches to diffeomorphism-invariant models using different techniques. That is, they show that one shouldn't expect observables or correlations to be differentiable functions and that spatiotemporal relations need to be considered part of the physical content of our models too.

**Declaration of competing interest**

The authors declare the following financial interests/personal relationships which may be considered as potential competing interests: Alvaro Mozota Frauca reports financial support was provided by Polytechnic University of Catalonia.

## Data availability

No data was used for the research described in the article.